\documentclass[11pt,a4paper]{article}
\usepackage{amsmath, amsthm} 
\usepackage{amsfonts}
\usepackage{amssymb,graphics,psfrag}
\usepackage[bottom]{footmisc}
\usepackage{hyperref}
\usepackage{mathrsfs}
\usepackage{cite}
\usepackage{setspace}
\usepackage{color}

\def\hybrid{\topmargin -10pt    \oddsidemargin 0pt
        \headheight 0pt \headsep 0pt
       \textwidth 6.25in       % A4 paper
      \textheight 9.5in       % A4 paper
        \marginparwidth .875in
        \parskip 5pt plus 1pt   \jot = 1.5ex}
\hybrid
\numberwithin{equation}{section}
\numberwithin{table}{section}
\begin{document}

\thispagestyle{empty}
\rightline{\small}
\vskip 3cm
\noindent
\begin{spacing}{1.1}
%\begin{onehalfspacing}
\noindent
{\LARGE \bf  The Algebra of Diffeomorphisms from the World Sheet
 }
%\end{onehalfspacing} 
 \end{spacing}
\vskip .1cm
\begin{center}
\linethickness{.06cm}
\line(1,0){447}
\end{center}
\vskip .8cm
\noindent
{\large \bf Waldemar Schulgin}

\vskip 0.2cm

{\em  \hskip -.05 cm Universit\'e Libre de Bruxelles and International  Solvay
Institutes}
\vskip -0.15cm
{\em \hskip -.05cm ULB-Campus Plaine CP231}
\vskip -0.15cm
{\em \hskip -.05cm B-1050 Brussels, Belgium}
\vskip -0.10cm
{\tt \hskip -.05cm waldemar.schulgin AT ulb.ac.be }
\vskip1cm

\noindent
{\large \bf Jan Troost}
\vskip 0.2cm
{\em \hskip -.05cm Laboratoire de Physique Th\'eorique\footnote{Unit\'e Mixte du CNRS et
    de l'Ecole Normale Sup\'erieure associ\'ee \`a l'universit\'e Pierre et
    Marie Curie 6, UMR
    8549.}}
    \vskip -.15cm
{\em \hskip -.05cm Ecole Normale Sup\'erieure}
 \vskip -.15cm
{\em \hskip -.05cm 24 rue Lhomond, 75005 Paris, France}

\vskip 1cm

\vskip0.6cm

\noindent {\sc Abstract:} 
The quantum theory of a massless spin two particle is 
strongly constrained by diffeomorphism invariance, which  is in turn implied
by unitarity.
We explicitly exhibit the space-time diffeomorphism algebra of string theory,
realizing it in terms of world sheet vertex operators. 
Viewing diffeomorphisms as field redefinitions in the two-dimensional
conformal field theory renders the calculation of their algebra
straightforward. Next, we generalize the analysis to combinations
of space-time anti-symmetric tensor gauge transformations and diffeomorphisms.
We also point out a left-right split of the algebra combined with a twist
that reproduces the C-bracket of double field theory. We further compare our 
derivation to an analysis in terms of marginal deformations as well as
 vertex operator algebras.

\newpage

\tableofcontents

\section{Introduction}
String theory is a theory containing a massless spin two
particle. There is a classic argument that a consistent interacting
theory of a massless spin two particle is necessarily a theory of
general relativity in the sense that the non-linear interactions are
constrained by the requirement of general coordinate invariance.  The
gauge symmetry must survive in the quantum theory to guarantee
unitarity through the decoupling of unphysical degrees of
freedom.\footnote{The history of this argument is reviewed in detail
  in the foreword to \cite{Feynman:1996kb} and a version of the
  argument is contained in the lectures \cite{Feynman:1996kb}.}  This
argument applies to perturbative string theory since it contains a massless spin two particle and is
unitary.
In the literature, string perturbation theory is more explicitly
demonstrated to be diffeomorphism invariant through a combination of two
arguments. It is shown that the transformation of the metric under a
diffeomorphism corresponds to adding a BRST exact state to the string
state at hand. Secondly, BRST exact states decouple from string
amplitudes through a generalized canceled propagator argument (based
on analyticity of the S-matrix).  And therefore, string perturbation
theory is diffeomorphism invariant.

We desire to go one step further in rendering this classic argument more
explicit in the context of string theory. We not only wish to identify the
relevant gauge transformations corresponding to infinitesimal coordinate
transformations in space-time, but also compute the algebra of diffeomorphisms
in terms of world sheet vertex operators. Indeed, the gauge algebra of string theory
is much larger than the diffeomorphism algebra (see e.g. \cite{Zwiebach:1992ie}).
It is interesting per se to isolate in this huge symmetry algebra the
gauge algebra governing gravitational interactions and the equivalence principle
(see also e.g. \cite{Ghoshal:1991pu}).

We have given a general motivation for attacking this problem.  We
are also inspired by the desire to realize asymptotic symmetry
algebras in string theory.  See
 for instance \cite{Giveon:1998ns,Troost:2011ud} and
\cite{Barnich:2010eb,Schulgin:2013xya,Strominger:2013jfa,Geyer:2014lca}
for older $AdS$ and more recent flat space applications of an understanding of
these asymptotic symmetries on the world sheet.  The
asymptotic symmetry algebra consist of asymptotically non-trivial
diffeomorphisms. A natural prerequisite to understanding their
world sheet counterparts is to understand the algebra of
diffeomorphisms on the world sheet in detail. A third motivation for our work
arises in the typically stringy extension of the algebra of
diffeomorphisms to also include anti-symmetric tensor gauge
transformations. The combination of the two naturally suggests a split
of these space-time gauge transformations in terms of world sheet left-
and right-movers. These may play a role
in our understanding of double field theory and T-duality (see e.g. \cite{Zwiebach:2011rg,Hohm:2013bwa} 
for reviews), with
potentially interesting applications (e.g. to non-geometric regions of 
string vacua, or to desingularizing cosmological evolutions).

The paper is organized as follows. In section \ref{difas}, we show how to interpret space-time
diffeomorphisms as field redefinitions in the two-dimensional conformal field theory. We demonstrate in the
covariant BRST formalism that the representation of these diffeomorphisms on the world sheet indeed
satisfies the expected diffeomorphism algebra. We then add anti-symmetric gauge transformations
to the algebra.
Their crucial property will be that exact shifts of the anti-symmetric tensor do not influence the world sheet
equations of motion, nor the world sheet energy-momentum tensor. That will allow us to explicitly
compute  the
mixed algebra of diffeomorphisms and anti-symmetric gauge transformations from the world sheet as well.

In section \ref{comparisons}, we compare our derivation to other approaches. In particular, we connect
to the approach in which deformations of the action are treated in conformal perturbation theory.
These approaches are related to the calculation of world sheet vertex operator algebras,
which mimic the diffeomorphism algebra at the classical level, but also
contain $\alpha'$ corrections 
\cite{Losev:2005pu,Hohm:2013jaa,Schulgin:2013xya}.

In the first appendix, we stress that, though operators $c \tilde{c} \, O^{matter}$  that we insert
in string correlation functions typically need to be of (matter) dimension $(1,1)$, this does not
need to hold for pure gauge operators, e.g. for diffeomorphism vertex operators. Those are automatically
on-shell.

In the second
 appendix \ref{doubled}, we point out  how to factor and twist
 the algebra of diffeomorphisms and anti-symmetric gauge 
transformations into an algebra dependent on world sheet left-movers and right-movers independently, 
and satisfying the C-bracket commutator.

\section{The algebra of gauge transformations}
\label{difas}
In this section, we derive the algebra of diffeomorphisms, and anti-symmetric gauge transformations
from a world sheet perspective.

\subsection{Diffeomorphisms in the path integral}
To implement diffeomorphisms in terms of the two-dimensional world
sheet theory on the fundamental string propagating in flat space, we
can turn to a path integral formalism for the gauge fixed world sheet
theory, which gives rise to a two-dimensional conformal field theory.  Ignoring world sheet
ghosts,  we find the path integral:
\begin{eqnarray}
Z &=& \int [ D X (z,\bar{z}) ] \exp \big( - S[X (z,\bar{z})] \big) 
\label{pathintegral}
\end{eqnarray}
where the world sheet action is
\begin{eqnarray}
S &=& \frac{1}{2 \pi \alpha'} \int d^2 z \, \eta_{\mu \nu}\,  \partial X^\mu \bar{\partial} X^\nu \, ,
\label{initialaction}
\end{eqnarray}
and where we can allow for operator insertions representing string scattering.
We wish to consider the effect of 
target space diffeomorphisms on the path integral. Diffeomorphisms in the target space
can be interpreted on the world sheet as field redefinitions:
\begin{eqnarray}
X^\mu(z,\bar{z}) & \rightarrow & X^\mu(z,\bar{z}) + \xi^\mu \big(X^\nu (z,\bar{z})\big) \, .
\label{initialfieldredef}
\end{eqnarray}
%v2c
When we perform the field redefinition on the action, we find that to first
order, the metric is shifted by the diffeomorphism according to the standard rule. 
%v2
The measure $[ D X]$ will transform unless we introduce the traditional measure factor $\sqrt{-G[X]}$
in equation (\ref{pathintegral}), where $G[X]$ is the determinant of the space-time metric $G_{\mu \nu}$.
Indeed, the measure factor will pick up a Jacobian transformation factor, while the square root
of the determinant will transform inversely under diffeomorphisms.
When we compute the commutator of gauge transformations, after a first application of a diffeomorphism,
the metric becomes non-trivially dependend on the coordinates $X$, and this measure factor will
become non-trivial.

%hepthv2 In a second application of a diffeomorphism then, this measure factor needs
%to be taken into account. 

If we concentrate on the world sheet kinetic term, and compute the algebra of diffeomorphisms,
we find the standard algebra 
%v2
-- see later for the details --:
\begin{eqnarray}
[ \xi^1 , \xi^2 ] &=& {\cal L}_{\xi^1} \xi^2 - {\cal L}_{\xi^2} \xi^1 \, ,
\label{diffeocommutator}
\end{eqnarray}
represented on the space of metric couplings in the two-dimensional field theory. To conclude that
we realize the standard diffeomorphisms, we must argue that the measure factor $\sqrt{-G}$ will not
lead to non-trivial contributions to string scattering. In  \cite{Evans:1989xq}, where the above
argument can also be found, it is argued
that this is the case because of the tracelessness of the graviton polarization tensor.
The same type of reasoning that we make above
would hold in curved spaces.
This is the underlying rationale for claiming that string theory in curved space will be diffeomorphism
invariant. Anti-symmetric gauge transformations can also be treated in this formal path integral argument.

There are several reasons to proceed beyond this. The first is that
one may wish to carefully regularize the path integral. A second
reason is that one wants to understand better the space of
deformations of the world sheet theory, and in particular distinguish
those that are trivial from the target space perspective from those
that are not. E.g. one would like to identify BRST exact deformations
from others in string theory, through a more direct analysis of the
shifts of the string state. Another reason is that one can have
diffeomorphisms that do not fall fast enough at infinity to make them
trivial field redefinitions. They can then generate a large and
interesting asymptotic symmetry group. One may then wish for an
explicit realization of the symmetry algebra on the world sheet in
terms of charge operators. Any of these motivations lead us to attempt
to analyze the above transformations in some more detail from a world
sheet perspective.

\subsection{Diffeomorphisms as field redefinitions}
We wish to be more explicit about how to compute the algebra of diffeomorphisms from a world sheet
perspective.
We start from a world sheet theory on a flat target space, with fields $X^\mu$ and metric $\eta_{\mu \nu}$
and action (\ref{initialaction}). The fields $X^\mu(z, \bar{z})$ are shifted as in the field
redefinition (\ref{initialfieldredef}).
The world sheet action after applying
the diffeomorphism becomes:
\begin{eqnarray}
S &=& \frac{1}{2 \pi \alpha'} \int d^2 z\,  \eta_{\mu \nu} \, \partial \left(X^\mu+\xi^\mu\right) \bar{\partial} \left(X^\nu+\xi^\nu\right)
\nonumber \\
& \approx &  \frac{1}{2 \pi \alpha'} \int d^2 z \, \left(\eta_{\mu \nu}+\partial_\mu \xi_\nu+ \partial_\nu \xi_\mu\right)\, 
 \partial X^\mu \bar{\partial} X^\nu
\label{shiftedaction1}
\end{eqnarray}
where we always must think of the shift $\xi^\mu(X^\nu)$ as field dependent. We work to leading order in the
expansion in the deformation $\xi$,
and in the above formula, we have used the Minkowksi metric to lower the index on $\xi$.
There are now at least 
 two useful perspectives on the shifted action. One that links quickly to the traditional
perspective on diffeomorphisms is to view:
\begin{eqnarray}
\delta_\xi S & = &  \frac{1}{2 \pi \alpha'} \int d^2 z \, \left(\partial_\mu \xi_\nu+ \partial_\nu \xi_\mu\right)
 \partial X^\mu \bar{\partial} X^\nu
\end{eqnarray}
as the integrated form of a vertex operator associated to a BRST exact
state in the original theory.\footnote{There is a subtlety here
  in that the integrand of this vertex operator is often assumed to be
  of world sheet dimension $(1,1)$, although there is no such
  restriction on $\xi$ from the target space point of view. We discuss
  this subtlety in appendix \ref{marginality}.} When we assume the integrand of the vertex
operator to be marginal, this perspective says that we have performed a 
marginal deformation of the world sheet theory that we could treat in conformal
perturbation theory. We will return to this perspective in section \ref{comparisons}.
 A second perspective
is to concentrate on the first line of (\ref{shiftedaction1}) and to
say that the new world sheet theory is very much like the old one,
except for the fact that the field $X^\mu$ has been replaced by the
field $X^\mu + \xi^\mu$. In the latter perspective, we draw
conclusions like the fact that the operator product expansion in the
deformed theory is (at $\alpha'=2$):
\begin{eqnarray}
(X + \xi)^\mu (z,\bar{z}) (X+\xi)^\nu(w,\bar{w}) & \approx & - \eta^{\mu \nu} \log |z-w|^2 \, ,
\label{defope}
\end{eqnarray}
and that the metric is still the initial metric $\eta_{\mu \nu}$, while
the energy-momentum tensor reads:
\begin{eqnarray}
T^m(\xi) &=& - \frac{1}{2} \, \eta_{\mu \nu} \, \partial \left(X+\xi\right)^\mu \partial \left(X+ \xi\right)^\nu \, .
\label{enmom}
\end{eqnarray}
In this perspective, in the BRST covariant treatment, the BRST charge
of the new theory is given in terms of the deformed  energy-momentum tensor (\ref{enmom}).
\subsection{The algebra of diffeomorphisms}
%v2
Let's now consider these transformations in the covariant BRST formalism.
In a first step, we have an undeformed BRST vertex operator:
\begin{equation}
Q_B(0)=\frac{1}{2\pi i}\oint \left(dz\, j_B-d\bar z\, \tilde j_B\right) \, ,
\end{equation}
where the ghost currents $j_B$ and $\tilde{j}_B$
depend on the undeformed world sheet energy-momentum tensor.
We perform a shift of the background by a BRST exact operator that is obtained by computing
the commutator of the undeformed BRST charge with a first seed\footnote{The seed is the state on  which we act with the BRST operator to obtain a BRST exact, or pure
gauge state. It can be thought off as the gauge parameter.}
operator $s_1$, depending
on a parameter $\xi^1$:
\begin{eqnarray}
s_1 &=& \tilde c\, \xi^1_\mu \bar\partial X^\mu -c\, \xi^1_\mu \partial X^\mu\, .
\label{firstseed}
\end{eqnarray}
When we compute the commutator, we find the BRST exact operator:
\begin{eqnarray}
q&=& [Q_B(0),s_1(\xi^1)] \nonumber \\
& = & \left(\partial_\mu \xi^1_\nu+\partial_\nu \xi^1_\mu \right)c\tilde c\, \partial X^\mu\bar\partial X^\nu\nonumber\\
%&&-\frac{\alpha'}{4}
%(\partial_\mu\partial^\mu \xi_\nu^L \ \partial c  \tilde{c} \ \bar\partial X^\nu 
%+ \partial_\mu\partial^\mu \xi_\nu^R \ c \bar{\partial} \tilde{c}\  \partial X^\nu)\nonumber \\
& &  + \mbox{other (left,right) ghost structures} \, .
%v2c added '.'
%\label{exactoperator}
\end{eqnarray}
The shift of the background by this operator shifts
%v2c
 the metric coupling of the background
according to the rule:
\begin{eqnarray}
\eta_{\mu \nu} & \rightarrow & \eta_{\mu \nu} +  \left(\partial_\mu \xi^1_\nu+\partial_\nu \xi^1_\mu \right)
\, ,
\end{eqnarray}
thus implementing the first diffeomorphism.
%
% we obtained after
%the first field redefinition, with parameter $\xi_1$.

To compute the commutator of two diffeomorphisms, we wish
to perform a second BRST exact shift of the background.
%\footnote{For the first, standard BRST exact
%shift, see appendix \ref{brstexactstate}.}
The BRST-charge after one deformation is given by
\begin{equation}
Q_B(\xi^1)=\frac{1}{2\pi i}\oint \left(dz\, j_B-d\bar z\, \tilde j_B\right) \, ,
\end{equation}
with matter energy-momentum tensor depending on $\xi^1$ as in equation (\ref{enmom})
\begin{eqnarray}
j_B&=&c\, T^{\rm{m}} (\xi^1) +:bc\partial c:+\frac{3}{2}\partial^2 c
\, .
\end{eqnarray}
The second seed
%
%\footnote{The seed is the state on  which we act with the BRST operator to obtain a BRST exact, or pure
%gauge state. It can be thought off as the gauge parameter.}\footnote{See also \cite{Ghoshal:1991pu} for an analysis
%of more generic seed states.}
that we wish to consider is most naturally written in terms of the deformed
fields $X^\mu + \xi^{1\mu}$, such that we find
%v2
-- this should be compared
to (\ref{firstseed}) --:
\begin{eqnarray}
s_2 &=& \tilde c\, \xi^{2} (X + \xi^1) \cdot\bar\partial (X+\xi^1) - c \, \xi^2(X+\xi^1) \cdot \partial (X+\xi^1)
\, .
\end{eqnarray}
The contraction is done with the flat background metric -- remember that we took the perspective in which
the metric is not shifted after the first deformation. The calculation of the BRST exact state that
results from this seed is identical to the standard calculation in flat space (thanks to
exact operator product expansions like (\ref{defope})), with the mere replacement
of the field $X$ by $X +\xi^1$, and we therefore find the result:
\begin{eqnarray}
C_{12} &=& \left[Q_B(\xi^1), s_2\right] 
\nonumber \\
&=& c \tilde{c}\,  \Big( \partial_\mu \xi^2_\nu \left(X + \xi^1\right) + \partial_\nu \xi^2_\mu \left(X + \xi^1\right) \Big)
\partial \left(X + \xi^1\right)^\mu \bar{\partial} \left(X+\xi^1\right)^\nu
\nonumber \\
& & +\,  \mbox{other ghost structures} \, .  
\end{eqnarray}
Finally, we are ready to compute the result of the commutator of two diffeomorphisms. We take
into account the linear shift in the background in $\xi^1$, and anti-symmetrize the above
expression in the deformations labeled $1$ and $2$, which results in the commutator:
\begin{eqnarray}
C &=& \left[ Q_B(0), s_1\right] + \left[Q_B(\xi^1),s_2)\right]  - ( 1 \leftrightarrow 2) 
\nonumber \\
&=&
 c \tilde{c} \, \Big( \partial_\mu 
\left(\xi^{1\rho} \partial_\rho \xi^2_\nu - \xi^{2\rho} \partial_\rho \xi^{1\nu}\right)
+ \partial_\nu 
\left(\xi^{1\rho} \partial_\rho \xi^2_\mu - \xi^{2\rho} \partial_\rho \xi^1_{\mu}\right)\Big)
\partial X^\mu  \bar{\partial} X^\nu 
\nonumber \\
& & +\,  \mbox{other ghost structures}\, .
\end{eqnarray}
This is the BRST exact shift of the background metric that results from consecutive gauge transformations
$1$ and $2$, minus $2$ then $1$.
Through expansions in the  shifts $\xi^1$ and $\xi^2$, we confirm the expected 
result (\ref{diffeocommutator}). With field redefinitions as our guide, the calculation of the commutator
in the two-dimensional quantum field theory has become
 a close analogue of its classical, geometric counterpart.

\label{diffsonly}

\subsection{The algebra including anti-symmetric gauge transformations}
Anti-symmetric gauge transformations in string theory are closely
related to diffeomorphisms. Under T-duality, the two types of
gauge symmetries become equivalent. It is therefore natural to attempt
to treat them on the same footing. Still, in the canonical treatment
they need separate care. Let's analyze how they modify the picture
drawn for diffeomorphisms in subsection \ref{diffsonly}. We can introduce anti-symmetric
gauge transformations as follows. In the first perturbation, we consider also a shift
of the action which is a total derivative on the world sheet :
\begin{eqnarray}
\delta_{\tilde{\xi}} S &=&- \frac{i}{2\pi \alpha'}\int   d\left(\tilde\xi_\mu \bar\partial X^\mu\, d\bar z+ \partial X^\mu \tilde\xi_\mu\,  d z\right)
\nonumber \\
&=& \frac{1}{2 \pi \alpha'} \int d^2 z \left(\partial \tilde{\xi}_\mu \bar{\partial} X^\mu
- \partial X^\mu \bar{\partial} \tilde{\xi}_\mu\right) \, .
\end{eqnarray}
where in the first line $d$ denotes the exterior derivative applied to the pull-back of the one-form 
$\tilde{\xi}$ to the world sheet.
Since this perturbation is a total derivative on the world sheet, it won't affect the world sheet
propagator, energy-momentum tensor and so on.
When we perform the analysis of the second shift, using the covariant BRST formalism, we
must also generalize the seed to include anti-symmetric tensor gauge transformations:
\begin{eqnarray}
s_2(\xi^2,\tilde{\xi}^2) &=&\tilde c\, \left(\xi^2+\tilde{\xi}^2\right) \cdot\bar\partial \left(X+\xi^1\right) - c \, \left(\xi^{2}-\tilde{\xi}^2\right) \cdot \partial \left(X+\xi^1\right)
\, .
\end{eqnarray}
The second BRST exact shift will then be equal to :
\begin{eqnarray}
\label{withgaugetransfo}
C_{12} &=& \left[Q_B(\xi^1), s_2(\xi^2,\tilde{\xi^2})\right] 
\nonumber \\
&=& c \tilde{c} \, \Big( \partial_\mu \left(\xi^2_\nu\left(X + \xi^1\right)+\tilde{\xi}^2_\nu\left(X + \xi^1\right)\right)  + \partial_\nu 
\left(\xi^2_\mu\left(X + \xi^1\right)-\tilde{\xi}^2_\mu\left(X + \xi^1\right)\right)  \Big)
\partial (X^\mu + \xi^{1\mu}) \bar{\partial} (X^\nu+\xi^{1\nu})
\nonumber \\
& & + \, \mbox{other ghost structures}  \, .
\end{eqnarray}
The commutator of two gauge transformations is equal to:
\begin{eqnarray}
C &=& \left[ Q_B(0), s_1 (\xi^1,\tilde{\xi}^1 ) \right] + \left[Q_B(\xi^1),s_2(\xi^2,\tilde{\xi^2}) \right]  - ( 1 \leftrightarrow 2) 
\nonumber \\
&=&
 c \tilde{c} \, \Big( \partial_\mu 
(\xi^{1\rho} \partial_\rho \xi^2_\nu - \xi^{2\rho} \partial_\rho \xi^{1\nu})
+ \partial_\nu 
(\xi^{1\rho} \partial_\rho \xi^2_\mu - \xi^{2\rho} \partial_\rho \xi^{1\mu})\Big)\ 
\partial X^\mu  \bar{\partial} X^\nu 
\nonumber \\
& &  +
 c \tilde{c} \, \Big( \partial_\mu 
(\xi^{1\rho} \partial_\rho \tilde{\xi}^2_\nu - \xi^{2\rho} \partial_\rho \tilde{\xi}^{1\nu})
- \partial_\nu 
(\xi^{1\rho} \partial_\rho \tilde{\xi}^2_\mu - \xi^{2\rho} \partial_\rho \tilde{\xi}^{1\mu})\Big)\ 
\partial X^\mu  \bar{\partial} X^\nu 
\nonumber \\
& &  +
 c \tilde{c}\,  \Big( -\partial_\mu \xi^{1\rho} \partial_\nu \tilde{\xi}^2_\rho
              +\partial_\nu \xi^{1\rho} \partial_\mu \tilde{\xi}^2_\rho
              +\partial_\mu \xi^{2\rho} \partial_\nu \tilde{\xi}^1_\rho
              -\partial_\nu \xi^{2\rho} \partial_\mu \tilde{\xi}^1_\rho\Big)\ 
\partial X^\mu  \bar{\partial} X^\nu 
\nonumber \\
& & + \, \mbox{other ghost structures}\, .
\end{eqnarray}
If we use the notations $D(\xi)$
for 
target space diffeomorphisms and $A(\xi)$ for anti-symmetric tensor gauge transformations
when they are represented on the metric and anti-symmetric tensor couplings in the world sheet
action,
we can summarize the algebra of these generators:\footnote{We have $({\cal
    L}_{\xi^1} \tilde{\xi}^2)_\mu = \xi^{1\rho} \partial_\rho
  \tilde{\xi}^2_{ \mu} + \partial_\mu \xi^{1\rho} \tilde{\xi}^2_{ \rho}$
  since $\tilde{\xi}$ is thought off as a one-form.}
\begin{eqnarray}
{[} D (\xi^1),D(\xi^2) ] &=& D([\xi^1,\xi^2]) \, , 
\nonumber \\
{[} A (\tilde{\xi}^1) ,A(\tilde{\xi}^2) ] &=& 0\, , 
\nonumber \\
{[} D (\xi^1),A(\tilde{\xi}^2) ] &=& A( {\cal L}_{\xi^1} \tilde{\xi}^2) \, .
\end{eqnarray}
We have the freedom to add a total derivative term to the argument of the anti-symmetric
tensor gauge transformation in the last line on the right hand side -- see e.g.  \cite{Zwiebach:2011rg} for a
clear  discussion -- to obtain the Courant bracket (with extra automorphism symmetry):
\begin{eqnarray}
{[} D (\xi^1),D(\xi^2) ] &=& D([\xi^1,\xi^2])\, ,  
\nonumber \\
{[} A (\tilde{\xi}^1) ,A(\tilde{\xi}^2) ] &=& 0\, , 
\nonumber \\
{[} D (\xi^1),A(\tilde{\xi}^2) ] &=& A( {\cal L}_{\xi^1} \tilde{\xi}^2 - \frac{1}{2} d (\xi^1 \cdot \tilde{\xi}^2)) \, .
\end{eqnarray}
In appendix \ref{doubled}, we demonstrate the existence of an algebra which is factorized in the
left- and right-movers on the world sheet. The algebra has a commutator equal to 
the C-bracket algebra of \cite{Siegel:1993th,Hull:2009zb,Zwiebach:2011rg}. 

\subsection{Contributions}
\label{contribs}
We want to make a brief remark on the terms that survive in the final result. For simplicity,
we concentrate on the diffeomorphism algebra. If we denote by $Y=X+\xi$ the redefined field,
and attach an index $Y$ to the quantities in the redefined frame, then we find the following
relation between the BRST exact state $q^Y$ in the $Y$ basis and the BRST exact states
$q^X$ in the $X$ basis:
\begin{eqnarray}
q^Y(\xi^2)&=&
q^X(\xi^2)+q^X(\xi^{1\rho}\partial_\rho\xi^2)+{\rm{  terms\ symmetric\ in\ }\xi^1\ {\rm{and}}\ \xi^2}
\, .  \label{qs}
\end{eqnarray}
The commutator of diffeomorphisms is then given by the usual algebra of vector fields, in
the original $X$ frame:
\begin{eqnarray}
q^X(\xi^1)+q^Y(\xi^2)-q^X(\xi^2)-q^Y(\xi^1)&=&q^X([\xi^1,\xi^2]) \, .
\end{eqnarray}
We see that only the second term in equation (\ref{qs}) survives in the end result.
Yet a third way to present the calculation is to rewrite the BRST exact term in the $Y$-frame in terms
of variations of the BRST charge, the seed, as well as the operator product expansion:
\begin{eqnarray}
q^Y(\xi^2)&=&[Q^Y_B,s^Y(\xi^2)]\Big|_{OPE^Y}  \\
&=&[Q^X_B+\delta_{\xi^1}Q_B^X, s^X(\xi^2)+\delta_{\xi^1}s^X(\xi^2)]\Big|_{OPE^X}+[Q^X_B+\delta_{\xi^1}Q_B^X, s^X(\xi^2)+\delta_{\xi^1}s^X(\xi^2)]\Big|_{\delta_{\xi^1} OPE^X}\nonumber\\
&=&[Q_B^X,s^X(\xi^2)]\Big|_{OPE^X}+[Q_B^X,\delta_{\xi^1}s^X(\xi^2)]\Big|_{OPE^X}+[\delta_{\xi^1}Q_B^X,s^X(\xi^2)]\Big|_{OPE^X}
+[Q_B^X,s^X(\xi^2)]\Big|_{\delta_{\xi^1}OPE^X} \nonumber 
\end{eqnarray}
We have for instance used the notation ${\delta_{\xi^1}OPE^X}$ to
indicate that after deformation, the operator product expansions for
the field $X$ have shifted, et cetera.  Once we consider the
commutator action of the diffeomorphisms all the terms except the
contribution from the second term will sum to zero. This is a 
 version of the observation in equation \eqref{qs} in conformal perturbation theory, namely that the
crucial term is a variation of the first diffeomorphism parameter with
respect to the second, and that other variations in the perturbation series drop out.

%%%%
\section{Comparison to other methods}
\label{comparisons}
In this section, we wish to  compare our explicit world sheet
identification of the diffeomorphism algebra inside the large gauge
algebra of string theory to various algebraic structures that have
been identified in string theory in the past. We indicate where these
structures relate to the algebra we have exhibited, and how they differ.
We permit ourselves to wander slightly, in order to overlap with
various approaches.
Our stroll illustrates that the smooth sailing we experienced in section 
\ref{difas} was due to our choice of route.

The literature has mostly concentrated on (physical or pure gauge) deformations
of the world sheet theory in terms of conformal perturbation theory.
We can fit an action deformed by a diffeomorphism in this category by
viewing the shift of the action under the field redefinition
$X \rightarrow X + \xi$ as a deformation of the original theory, and 
performing perturbation theory for the correlators:
\begin{eqnarray}
\langle \dots \rangle_\xi &=& \sum_{n=0}^\infty  \left\langle 
\frac{1}{n!}  \left(\frac{1}{2 \pi \alpha'} \int d^2 z\, \big( \partial \xi\cdot \bar{\partial}X
+ \partial X\cdot \bar{\partial} \xi\big) \right)^n \dots \right\rangle_0 \, .
\label{deformed}
\end{eqnarray}
All correlators in the perturbed theory obtain contributions from the
extra insertions. The technique we used in section \ref{difas} boils
down to an elementary, effortless resummation of conformal
perturbation theory, available only in the case where this deformation
is equivalent to a field redefinition. If one performs several
consecutive deformations, one can ask whether the algebra of deforming
vertex operators (as in equation (\ref{deformed})) is related to the
diffeomorphism algebra we found in section \ref{difas}. We will
address this question in this section, and connect it to various
approaches in the literature.

\subsection{The chiral cohomological Gerstenhaber algebra}
In \cite{Lian:1992mn}, BRST algebraic structures in chiral or
open bosonic string theory were discussed. In particular, the
compatibility between holomorphic normal ordering, the $b_{-1}$ curly
bracket operation and the chiral BRST cohomology were analyzed (see \cite{Lian:1992mn} for details). It was
shown that they form a Gerstenhaber algebra in cohomology. We can
read the relevant proofs of these properties slightly differently,
namely, in the full covariant Hilbert space.  The proofs of
\cite{Lian:1992mn} then say that if two states are BRST exact, their
curly bracket gives again an exact state, and the seed can be explicitly
calculated. Similarly, if three states are BRST exact, their curly bracket
Jacobiator is exact, and the seed of the Jacobiator can be explicitly
 calculated. These are interesting properties of the chiral gauge algebra in
string theory. They are closely related to properties of (chiral) vertex operator
algebras reviewed for instance in \cite{Kac:1996wd} and exploited
for chiral analogues of generalized diffeomorphism vertex operator algebras
in \cite{Losev:2005pu,Hohm:2013jaa,Schulgin:2013xya}.
However, to usefully compare to our results, we need to study a 
 non-chiral version of these algebras.

\subsection{The non-chiral Gerstenhaber algebra}
The curly bracket operation and its relation to the BRST operator, and marginal deformations
has been further extended and exploited in the non-chiral context, e.g. in 
\cite{Zeitlin:2009hc}. The deformation of a matter conformal field theory by a marginal
operator was written in 
\cite{Zeitlin:2009hc} in the language of the algebraic structures of  \cite{Lian:1992mn}, generalized to 
the non-chiral setting. In particular, this requires the introduction of a non-chiral form of normal ordering
which was chosen to be:
\begin{eqnarray}
NO (V_1,V_2) (0) &=& P_0 ( V_1(\epsilon) V_2(0)) \label{NO}
\end{eqnarray}
where $P_0$ is a projection operator on the term in the operator product expansion between $V_1$ and $V_2$ that
is independent of $\epsilon$, including its phase. Effectively, the normal ordering is taken to project
onto the non-singular term in both the holomorphic and the anti-holomorphic factor separately.
 The non-chiral curly bracket operation acts
as a chiral curly bracket on one side, times a normal ordering operation on the other. One then
again identifies the resulting algebra as a Gerstenhaber algebra.
(See  \cite{Lian:1992mn} and \cite{Zeitlin:2009hc} for details.)
For a recent analysis of how one recuperates diffeomorphisms and anti-symmetric gauge
transformations within this framework, see \cite{Zeitlin:2014xma}.

\subsection{Marginal perturbations}
An alternative way to link some of these interesting
algebraic structures to our elementary treatment is the following. We need to suppose
henceforth that the diffeomorphism operator by which we deform the action
is marginal.\footnote{We remind the reader that this is not generic, as reviewed
in appendix \ref{marginality}.} A perturbation of the action with the integral of a marginal $(1,1)$ vertex
operator $W^{(1,1)}$ corresponds to a deformation of the BRST charge of the form:
\begin{eqnarray}
Q^{def}_B(z) (\cdot) &=& Q_B \, \cdot + \{ W, \cdot \} \, , 
\nonumber 
\end{eqnarray}
where $W^{(1,1)}$ the doubly descended form of the operator $W=c \tilde{c} \, W^{(1,1)}$.
We have introduced the curly bracket operation $\{ .\, ,. \}$ which underlies the Gerstenhaber algebra.\footnote{Since
we hardly need any of its properties, we will not fully review this operation \cite{Lian:1992mn,Zeitlin:2009hc}.}
If we apply this knowledge to a diffeomorphism/anti-symmetric gauge transformation of the form:
\begin{eqnarray}
W^{(1,1)} &=& c \tilde c\ \Big( \partial \xi^L \cdot \bar \partial X+ \bar\partial \xi^R \cdot\partial X\Big) \, ,
\end{eqnarray}
then
we find the first order deformation:
\begin{eqnarray}
Q^{def}_B  &=&Q_B +\Big\{c\tilde c\ \Big( \partial \xi^L \cdot \bar \partial X+ \bar\partial \xi^R \cdot\partial X\Big),\cdot\Big\} \, .
\end{eqnarray}
The curly bracket replaces the first entry by the operator
corresponding to acting by descent once on the operator $c \tilde{c}\, 
W^{(1,1)}$, and takes a small contour integral of this operator around the one
it is acting on, with the $\epsilon$ regularization prescription
described of equation (\ref{NO}) (see \cite{Zeitlin:2009hc}). We therefore have:
\begin{eqnarray}
Q^{def}_B   &=&Q_B + \frac{1}{2 \pi i} \oint_{C_\epsilon}
(dz\,  c - d \bar{z}\,  \tilde{c}) \Big( \partial \xi^L \cdot \bar \partial X+ \bar\partial \xi^R \cdot\partial X\Big) \, .
\label{LZZ}
\end{eqnarray}
{From} the result (\ref{LZZ}), we conclude that the description of marginal deformations in the
formalism of \cite{Lian:1992mn,Zeitlin:2009hc} is
equivalent to the analysis of the influence of marginal deformations
on the BRST charge performed  in \cite{Sen:1990hh}.

The analysis of \cite{Sen:1990hh} is based on a
regularization procedure in conformal field theory that cuts out little disks from the integral
over the world sheet of the marginal deformation of the theory whenever
one encounters another vertex operator insertion.  The coordinate
independence of the deformation of the BRST charge has been analyzed
in \cite{Campbell:1990dz}. That analysis also showed the equivalence between the deformation of  \cite{Sen:1990hh}
and  the equal-time contour prescription of the deformation of the world sheet
energy-momentum tensor proposed in 
\cite{Evans:1989xq,Ovrut:1990ue}. Further developments can amongst
others be found in \cite{Ranganathan:1992nb,Ranganathan:1993vj,Cederwall:1995nc}.

The deformation of the BRST charge is obtained 
by analyzing the conformal Ward identities in the perturbed theory \cite{Sen:1990hh},
or by analyzing how conformal transformations influence the procedure of cutting out
disks around other vertex operator insertions \cite{Campbell:1990dz}. Our method in 
section \ref{difas}  directly resums conformal
perturbation theory. Of course, the methods of  \cite{Sen:1990hh,Campbell:1990dz}
were designed to be more general, and to apply to physical deformations.

Now that we have reviewed how the deformation of the BRST charge under
marginal deformations is coded in a deformation of the world sheet
stress-energy tensor of the equal-time contour form discussed
in\cite{Evans:1989xq,Ovrut:1990ue} we can apply the result of
\cite{Evans:1989xq} where the deformation of the energy-momentum
tensor is interpreted as a change in the background fields, e.g. the
metric.\footnote{See also \cite{Sen:1990hh} for a analysis of how deformations of the BRST charges
due to marginal deformations and due to shifts in the string field are related.}
  For a given deformation of the background metric $h_{\mu \nu}$,
we have the relation \cite{Evans:1989xq}:
\begin{eqnarray}\label{eqOV}
T (h_{\mu \nu}) +i[\Lambda(\xi),T(h_{\mu \nu})]=T (h_{\mu \nu} + \delta_\xi h_{\mu \nu}) \, 
\end{eqnarray}
between the shifted energy-momenum tensor, the vertex operator $\Lambda(\xi)$ corresponding
to the diffeomorphism, and the resulting perturbed energy-momentum tensor.
The commutator in equation (\ref{eqOV})  is an equal time commutator.
The generator of diffeomorphisms $\Lambda(\xi)$ was proposed to be the vertex operator:
 \begin{equation}\label{Lambda}
\Lambda(\xi)=\frac{1}{2\pi i\alpha'}\int\, d\bar z\, \xi_\mu(z,\bar z)\bar \partial X^\mu-\frac{1}{2\pi i\alpha'} \int \, dz\, \xi_\mu(z,\bar z)\partial X^\mu\end{equation}
with integrands of dimension $(1,0)$ and $(0,1)$, which imposes the conditions:
\begin{eqnarray}
\partial_\mu\xi^\mu=0\ , \qquad \partial_\mu\partial^\mu\xi^\nu=0 \ .
\end{eqnarray}
Now consider an undeformed background with $T(0)=-\frac{1}{2}
\eta_{\mu \nu} \partial X^\mu \partial X^\nu$
and shift the stress-energy tensor by transformations generated by $\Lambda ({\xi^1})$ and $\Lambda ({\xi^2})$.
With a slight abuse of notation, we find:
\begin{eqnarray}
\delta_{\xi^2}\delta_{\xi^1}T(0) &=&T({\xi^1})+i[\Lambda({\xi^2}),T({\xi^1})]\nonumber\\
&=& T(0)+i[\Lambda({\xi^1}),T(0)]+i[\Lambda({\xi^2}),T(0) ]- [\Lambda({\xi^2}), [\Lambda({\xi^1}),T(0)]] \, .
\end{eqnarray}
The commutator is then equal to:
\begin{eqnarray}\label{Trel}
\Big(\delta_{\xi^2}\delta_{\xi^1}-\delta_{\xi^1}\delta_{\xi^2}\Big)T(0)
&=&[T(0),\, [\Lambda({\xi^2}),\Lambda({\xi^1})]] \, .
\end{eqnarray}
To obtain this result we used the Jacobi identity for the energy-momentum tensor and the
operators $\Lambda(\xi^i)$.
In \cite{Schulgin:2013xya} we computed the vertex operator algebra between $\Lambda({\xi^1})$ and 
$\Lambda({\xi^2})$  
\begin{eqnarray}\label{comm}
{[} \Lambda({\xi^1}) , \Lambda({\xi^2}) {]}
&=&  
\Lambda \left(   
     \xi^{1 \rho} \partial_\rho \xi^{2}_\mu -\xi^{2 \rho} \partial_\rho \xi^{1}_\mu
 - \frac{1}{\alpha'}\,\left( 
\partial_\rho \xi^{1 \nu}  \, \partial_\mu \partial_\nu \xi^{2 \rho}-
\partial_\rho \xi^{2 \nu}  \, \partial_\mu \partial_\nu \xi^{1 \rho}\right)
\right) \, .
\end{eqnarray}
The equal time commutator between the $\Lambda(\xi^i)$ 
generates the target space diffeomorphism algebra at
leading order in $\alpha'$. 
The vertex operator algebra of the  $\Lambda(\xi^i)$ operators does
have $\alpha'$ corrections \cite{Schulgin:2013xya},  making the map between the diffeomorphism
algebra and the vertex operator algebra an isomorphism only
classically.\footnote{An alternative
set of $\alpha'$ corrections to the diffeomorphism and anti-symmetric gauge transformation algebra
was identified in \cite{Hohm:2013jaa,Hohm:2014eba} on the basis of a chiral vertex operator algebra
\cite{Losev:2005pu} in the doubled field theory formalism. These $\alpha'$ corrections are consistent with the
Green-Schwarz mechanism in heterotic string theory.} 
{From} this perspective a proper map of the diffeomorphism
action from the target space into the world sheet will include
$\alpha'$-corrections, namely, it is  more subtle than the map we used
in equation \eqref{Lambda}. 
It is important to note that the analysis of the vertex operator algebra does not take into account
effects such as the deformation of one seed by the other, nor does it take into account possible
corrections to the operator product relations after a single deformation (see subsection \ref{contribs}). 
Rather, the  vertex operator algebra
analysis should be viewed as taking place within conformal perturbation theory.

\section{Conclusions}
String theory is a unitary perturbative theory of quantum gravity. As
such we expect it to be diffeomorphism invariant on the basis of 
classic arguments \cite{Feynman:1996kb}.  We have explicitly exhibited the
diffeomorphism algebra of string theory through the interpretation of target space diffeomorphisms as
world sheet field redefinitions. 
%v2
 This allowed us to efficiently derive the space-time
algebra from the world sheet quantum field theory defining perturbative
string theory.
%v2 
The derivation of the full diffeomorphism algebra in a fully regularized setting 
%hepthv2 is new, and 
allows
to demonstrate explicitly that there are no $\alpha'$ corrections in the gauge algebra we identified.
Thus we have exhibited
%v2c
 the canonical, undeformed diffeomorphism algebra in the quantum world sheet
string theory. 
%v2
 We also compared our derivation to
algebraic structures in the literature, and showed the subtlety of the
comparison to vertex operator algebras which do exhibit $\alpha'$ corrections.

We believe these steps are useful in the enterprise of understanding
better asymptotic symmetry algebras in string theory, consequences of
diffeomorphisms for graviton scattering amplitudes, constraints
on doubled effective field theories, et cetera. Our set-up, suitably regularized,
is also a good template for similar
analyses in curved space, including asymptotic symmetry algebras in
asymptotically anti-de Sitter, linear dilaton, de Sitter, or less symmetric space-times.

\section*{Acknowledgments}
We would like to thank Costas Bachas, Glenn Barnich, Ben Craps, Giusseppe Policastro,
Ashoke Sen and Anton Zeitlin for useful discussions and correspondence.
The work of WS was partially supported by the ERC Advanced Grant "SyDuGraM", by a Marina Solvay fellowhip, by IISN-Belgium (convention 4.4514.08) and by the ``Communaut\'e Fran\c{c}aise de Belgique" through the ARC program.

%%%%%%%%%%\%\
\appendix

\section{Exactness, on-shellness and marginality}
\label{marginality}
Often in the string theory literature, on-shellness of a vertex operator and its marginality
are viewed as synonyms. Indeed, a matter vertex operator integrand should be marginal in order
for the integrated vertex operator to be conformally invariant. A related fact is
that a primary operator of the form $c \tilde{c} \, O_{matter}$ is BRST closed only if the matter
operator $O_{matter}$ is of dimensions $(1,1)$.

However, there is a slightly more general concept of on-shellness
which is important in the context of pure gauge degrees of
freedom. Indeed, we can define an operator to be on-shell if is it
BRST closed.  It is clear then that any  world sheet vertex
operator that is BRST exact is BRST closed, and therefore on-shell. No condition on its dimension or primary
nature are necessary. In other words, pure gauge degrees of freedom are always on-shell. This is logical in the sense that
we don't expect pure gauge degrees of freedom to be subject to a physical equation of motion (imposed by the requirement
of having a particular conformal dimension). (See \cite{Zwiebach:1992ie} for how these facts play out in the
larger context of
string field theory.)

Since this feature of pure gauge degrees of freedom is often set
aside, we wish to treat it in some detail in this appendix. We
illustrate this feature with an elementary example, in particular as it
pertains to the diffeomorphisms and anti-symmetric gauge
transformations discussed in the bulk of the paper.

\subsection{Diffeomorphisms and anti-symmetric gauge transformations}
\label{brstexactstate}
Let's  consider a flat target space with the standard BRST operator, energy-momentum tensor,
ghost system, et cetera \cite{Pol1}. 
%v2
We consider a seed of the form\footnote{See also \cite{Ghoshal:1991pu} for an analysis
of more generic seed states.}:
\begin{eqnarray}
s &=& \tilde c\, \xi^L_\mu \bar\partial X^\mu -c\, \xi^R_\mu \partial X^\mu\, .
\end{eqnarray}
Acting with the BRST operator gives rise to the BRST exact state:
\begin{eqnarray}
q&=&\left(\partial_\mu \xi^L_\nu+\partial_\nu \xi^R_\mu \right)c\tilde c\, \partial X^\mu\bar\partial X^\nu\nonumber\\
&&-\frac{\alpha'}{4}
(\partial_\mu\partial^\mu \xi_\nu^L \ \partial c  \tilde{c} \ \bar\partial X^\nu 
+ \partial_\mu\partial^\mu \xi_\nu^R \ c \bar{\partial} \tilde{c}\  \partial X^\nu)\nonumber \\
& &  + \mbox{other (left,right) ghost numbers}
\label{exactoperator}
\end{eqnarray}
where we dropped terms that are not of ghost numbers $(1,1)$. Of course, one can act on the full vertex
operator (\ref{exactoperator}) once more to demonstrate it is BRST closed without having to impose any conditions
on the parameters $(\xi^L,\xi^R)$.

\subsection{An example three-point function}
Let's turn to an example in which we can explicitly see how a non-marginal matter part of a vertex operator leads to a well-defined 
result for the pure gauge correlator, namely zero.
We show that a BRST exact massless vertex operator decouples from a three-point function in string
theory, through explicit computation. A standard technique to demonstrate this fact is  to use the contour argument, that has the BRST charge act on the other
two insertions of the three-point function (see e.g. \cite{Pol1}). We find the explicit calculation to be informative as well.
We calculate the three-point function of a Fourier mode of the  BRST exact operator $q$ of equation (\ref{exactoperator}) 
\begin{eqnarray}
q(k_1) &=& \left( i k^1_\mu e_\nu^L(k_1) + ik^1_\nu e_\mu^R(k_1) \right) c\tilde c\, \partial X^\mu\bar\partial X^\nu\nonumber\\
&&+\frac{\alpha'}{4}
k_1^2 \left(e_\mu^L \,  \partial c  \tilde{c}\,  \bar\partial X^\mu 
+ e_\mu^R\,  c \bar{\partial} \tilde{c} \, \partial X^\mu\right)\nonumber \\
& &  + \mbox{other (left,right) ghost numbers}
\end{eqnarray}
with two physical tachyon vertex operator insertions -- we follow the
conventions of \cite{Pol1} closely:
\begin{eqnarray}
S_3 &=& \left\langle  : q(k_1) (z_1,\bar{z}_1) : \ : c \tilde{c} \, e^{i k_2 \cdot X (z_2,\bar{z}_2)} :\ : c\tilde c\, e^{i k_3 \cdot X (z_3,\bar{z}_3)} : \right\rangle
\nonumber \\
&=&  |z_{12} \, z_{13}\,  z_{23}|^2 \ |z_{12}|^{\alpha' k_1 \cdot k_2}\ |z_{13}|^{\alpha' k_1 \cdot k_3}\ |z_{23}|^{\alpha' k_2 \cdot k_3} 
\nonumber \\
& \times & \left[
\left( i k^1_\mu e_\nu^L + ik^1_\nu e_\mu^R\right) \left(-i \frac{{\alpha'}}{2}\right)^2
\left(\frac{k_2^\mu}{z_1-z_2} + \frac{k_3^\mu}{z_1-z_3}\right)\left(\frac{k_2^\nu}{\bar z_1- \bar z_2} + \frac{k_3^\nu}{\bar z_1- \bar z_3}\right)\right.
\nonumber \\
& & +\left( - i \frac{\alpha'}{2}\right) \frac{\alpha' k_1^2}{4}  e_\mu^L \left( \frac{1}{z_1-z_2} + \frac{1}{z_1-z_3} \right) \left(\frac{k_2^\mu}{\bar z_1- \bar z_2} + \frac{k_3^\mu}{\bar z_1- \bar z_3}\right)
\nonumber \\
& &\left. +\left(- i \frac{\alpha'}{2}\right)   \frac{\alpha' k_1^2}{4} 
\left(\frac{k_2^\mu}{z_1-z_2} + \frac{k_3^\mu}{z_1-z_3}\right) e_\mu^R \left( \frac{1}{\bar z_1 - \bar z_2} + \frac{1}{ \bar z_1- \bar z_3} \right)
\right] \, .
\end{eqnarray}
We left out  the overall normalization factor (equal to $2 i \, g_{cl}^3\,  C_{S^2} / \alpha'$) as well as the overall delta-function 
corresponding to momentum conservation $(2 \pi)^{26} \ \delta^{26}\left(\sum_{i=1}^3 k^\mu_i\right)$. We now use momentum conservation to determine the overall
factor for both the $e_L$ and $e_R$ dependence:
\begin{eqnarray}
\alpha' \left( 2 k_1 \cdot k_2 + k_1^2\right) &=& \alpha' \left(k_3^2-k_2^2\right) = 0
\end{eqnarray}
where in the last equality, we exploited that the two tachyon vertex operators are on-shell and have equal mass. 
The amplitude, which at first sight looked like a gauge dependent off-shell quantity (because
of the factor with explicit dependence on the insertion points $z_{i}$), upon closer inspection is equal to zero. Both in the contour argument and in the explicit calculation, there
is no constraint on the momentum $k_1$ of the pure gauge vertex operator.  The ghost structure of the vertex operator is of primordial importance in both arguments. Our calculation is an elementary illustration of the fact that BRST closedness (or in 
this case, exactness) is
sufficient for a vertex operator to be on-shell. This fact is important in the bulk of the paper since we
wish to work with arbitrary gauge parameters $\xi$. Our story here is but a close-up of a more systematic, general,
and abstract
treatment within for instance closed
string field theory \cite{Zwiebach:1992ie}.

\section{Factorization and Doubling}
\label{doubled}
T-duality exchanges metric and anti-symmetric tensor components, and
exchanges diffeomorphisms for anti-symmetric gauge transformations.
In section \ref{difas}, the two types of transformations were on a
different footing. T-duality (see e.g. \cite{Giveon:1994fu}) feeds the desire to treat them
democratically. In order to do so, it is convenient to double the
number of space-time coordinates (see e.g. \cite{Zwiebach:2011rg} for an introduction to the
relevant concepts). We wish to introduce a left and a
right space-time coordinate $X^L$ and $X^R$, as well as their shifts
$\xi^L$ and $\xi^R$. There are now a number of different ways to proceed.

We wish to entertain the possibility to obtain an $O(d,d)$ invariant algebra as a result
of our procedure. To do so,  we can make the guess that the second seed on which our deformed
BRST charge acts should  be of the form:
\begin{eqnarray} 
s_2&=& 2\,  c\,  \xi^{L2} \cdot \partial X - 2\,  \tilde{c}\,  \xi^{R2} \cdot \bar{\partial} X
\, .
\label{twistedsecond}
\end{eqnarray}
Note how this is in contrast to what we did in section \ref{difas}. We have flipped the left
and right component of the second seed (only). This puts a particular twist on 
our gauge parameters. This idea leads to the following proposal for the BRST exact
shift of the background after performing transformations $1$ and $2$ consecutively:
\begin{eqnarray}
C_{12} &=& 2 c \tilde{c} \, \Big( \bar{\partial} \xi^{2L}(X^L+\xi^{1L},X^R+\xi^{1R}) \cdot {\partial} (X+ \xi^{1L})
+ {\partial} \xi^{2R}(X^L+\xi^{1L},X^R+\xi^{1R}) \cdot \bar{ \partial} (X+\xi^{1R}) \Big) \, . \nonumber\\
\end{eqnarray}
Given this starting point, it is straightforward to
 calculate the commutator. We expand to second order in $\xi$, using amongst
others the rule:
\begin{eqnarray}
\xi^{L2}\left(X^L+\xi^{L1},X^R+\xi^{R1}\right)\approx \xi^{L2}
+\xi^{L1\rho}\partial^L_\rho\xi^{L2}
+\xi^{R1\rho}\partial^R_\rho\xi^{L2} + \dots 
\end{eqnarray}
The full commutator can be summarized as:
\begin{eqnarray}
\label{CLR}
C &=&
2 c\tilde c\, 
\Big[-\partial\xi^{L2}\cdot \bar\partial \xi^{L1}+\partial \xi^{L1}\cdot \bar\partial \xi^{L2}-\bar \partial \xi^{R2}\cdot \partial \xi^{R1}+\bar \partial\xi^{R1}\cdot \partial \xi^{R2}\nonumber\\
&&+\bar\partial\Big(\xi^{L1\rho}\partial_\rho^L\xi^{L2}-\xi^{L2\rho}\partial_\rho^L\xi^{L1}+\xi^{R1\rho}\partial_\rho^R\xi^{L2}-\xi^{R2\rho}\partial_\rho^R\xi^{L1}\Big)\cdot \partial X\nonumber\\
&&+ \partial\Big(\xi^{L1\rho}\partial_\rho^L\xi^{R2}-\xi^{L2\rho}\partial_\rho^L\xi^{R1}+\xi^{R1\rho}\partial_\rho^R\xi^{R2}-\xi^{R2\rho}\partial_\rho^R\xi^{R1}\Big)\cdot \bar\partial X\Big]\, . \nonumber
\end{eqnarray}
We wish to express the commutator in terms of the gauge parameters. To
that end, we need to decide how we will interpret the first line in
(\ref{CLR}). We rewrite the first two terms in the first line as:
\begin{eqnarray}
C_{L1} &=& -c \tilde{c} \Big( \partial \left(\xi^{L2} \cdot \bar{\partial} \xi^{L1}\right)
- \bar{\partial} \left(\xi^{L2} \cdot \partial \xi^{L1}\right)
+ \bar{\partial} \left(\partial \xi^{L2} \cdot \xi^{L1}\right)
- \partial \left( \xi^{L1} \cdot \bar{\partial} \xi^{L2}\right)\Big)
\nonumber \\
&=&-
 c \tilde{c} \Big(\partial \left( \xi^{L2} \cdot \partial_\mu^R  \xi^{L1} \bar{\partial} X^\mu\right)
- \bar{\partial} \left(\xi^{L2} \cdot \partial_\mu^L \xi^{L1} \partial X^\mu\right)
+ \bar{\partial} \left(\partial_\mu^L \xi^{L2} \cdot \xi^{L1} \partial X^\mu\right)
- \partial \left( \xi^{L1} \cdot \partial_\mu^R \xi^{L2} \bar{\partial} X^\mu\right)\Big) \nonumber \, .
\end{eqnarray}
Using this rewriting, we find the commutators of gauge parameters:
\begin{eqnarray}
[ (\xi_{L1},0),(\xi_{L2},0)]
&=& 2  \left(\xi^{L1\rho}\partial_\rho^L\xi^{L2}-\xi^{L2\rho}\partial_\rho^L\xi^{L1},0\right)
\nonumber \\
& & + \left( \xi^{L2} \cdot \partial_\mu^L  \xi^{L1} - \xi^{L1} \cdot \partial_\mu^L \xi^{L2}
,-\xi^{L2} \cdot \partial_\mu^R  \xi^{L1} + \xi^{L1} \cdot \partial_\mu^R \xi^{L2} \right)\, ,
\nonumber \\
{[} (\xi_{L1},0),(0,\xi_{R2}) ]
&=&2 \left( - \xi^{R 2 \rho} \partial_\rho^R \xi^{L1}, \xi^{L1 \rho} \partial_\rho^L \xi^{R2}\right)
\, .
\label{algofpar}
\end{eqnarray}
On the right hand side, we read off the left parameter from the $\partial X$ terms, 
in accordance with the twist implemented in the second seed (\ref{twistedsecond}).
We have succeeded in twisting our construction such that
the resulting algebra gives rise to the $O(d,d)$ invariant C-bracket
\cite{Siegel:1993th,Hull:2009zb,Zwiebach:2011rg}.
This bracket also has a world sheet vertex operator algebra counterpart 
which exibits $\alpha'$ corrections found in  \cite{Schulgin:2013xya}.

We stress that we have improvised at the level of a twist on the 
space of gauge parameters, and the proposal for the resulting BRST exact state.
It would be interesting to find a good rationale behind the construction of the C-bracket
in this BRST covariant context.


\begin{thebibliography}{99}

%\cite{Feynman:1996kb}
\bibitem{Feynman:1996kb}
  R.~P.~Feynman, F.~B.~Morinigo, W.~G.~Wagner and B.~Hatfield,
  ``Feynman lectures on gravitation,''
  Reading, USA: Addison-Wesley (1995) 232 p. Foreword by
J.~Preskill and K.~S.~Thorne.

%\cite{Zwiebach:1992ie}
\bibitem{Zwiebach:1992ie}
  B.~Zwiebach,
  ``Closed string field theory: Quantum action and the B-V master equation,''
  Nucl.\ Phys.\ B {\bf 390} (1993) 33
  [hep-th/9206084].
  %%CITATION = HEP-TH/9206084;%%
 

%\cite{Ghoshal:1991pu}
\bibitem{Ghoshal:1991pu}
  D.~Ghoshal and A.~Sen,
  ``Gauge and general coordinate invariance in nonpolynomial closed string theory,''
  Nucl.\ Phys.\ B {\bf 380} (1992) 103
  [hep-th/9110038].
  %%CITATION = HEP-TH/9110038;%%
 

%\cite{Giveon:1998ns}
\bibitem{Giveon:1998ns}
  A.~Giveon, D.~Kutasov and N.~Seiberg,
 ``Comments on string theory on AdS(3),''
  Adv.\ Theor.\ Math.\ Phys.\  {\bf 2} (1998) 733
  [hep-th/9806194].
  %%CITATION = HEP-TH/9806194;%%
 

%\cite{Troost:2011ud}
\bibitem{Troost:2011ud}
  J.~Troost,
  ``The $AdS_3$ central charge in string theory,''
  Phys.\ Lett.\ B {\bf 705} (2011) 260
  [arXiv:1109.1923 [hep-th]].
  %%CITATION = ARXIV:1109.1923;%%
 
%\cite{Barnich:2010eb}
\bibitem{Barnich:2010eb}
  G.~Barnich and C.~Troessaert,
  ``Aspects of the BMS/CFT correspondence,''
  JHEP {\bf 1005} (2010) 062
  [arXiv:1001.1541 [hep-th]].
  %%CITATION = ARXIV:1001.1541;%%
 

%\cite{Schulgin:2013xya}
\bibitem{Schulgin:2013xya}
  W.~Schulgin and J.~Troost,
  ``Asymptotic symmetry groups and operator algebras,''
  JHEP {\bf 1309} (2013) 135
  [arXiv:1307.3423].
  %%CITATION = ARXIV:1307.3423;%%


%\cite{Strominger:2013jfa}\cite{Geyer:2014lca}
\bibitem{Strominger:2013jfa}
  A.~Strominger,
  ``On BMS Invariance of Gravitational Scattering,''
  arXiv:1312.2229 [hep-th].
  %%CITATION = ARXIV:1312.2229;%%




%\cite{Geyer:2014lca}
\bibitem{Geyer:2014lca}
  Y.~Geyer, A.~E.~Lipstein and L.~Mason,
  ``Ambitwistor strings at null infinity and subleading soft limits,''
  arXiv:1406.1462 [hep-th].
  %%CITATION = ARXIV:1406.1462;%%


\bibitem{Zwiebach:2011rg}
  B.~Zwiebach,
  ``Double Field Theory, T-Duality, and Courant Brackets,''
  Lect.\ Notes Phys.\  {\bf 851} (2012) 265
  [arXiv:1109.1782 [hep-th]].
  %%CITATION = ARXIV:1109.1782;%%


%\cite{Hohm:2013bwa}
\bibitem{Hohm:2013bwa}
  O.~Hohm, D.~L\"ust and B.~Zwiebach,
  ``The Spacetime of Double Field Theory: Review, Remarks, and Outlook,''
  Fortsch.\ Phys.\  {\bf 61} (2013) 926
  [arXiv:1309.2977 [hep-th]].
  %%CITATION = ARXIV:1309.2977;%%


%\cite{Losev:2005pu}\cite{Hohm:2013jaa}
\bibitem{Losev:2005pu}
  A.~S.~Losev, A.~Marshakov and A.~M.~Zeitlin,
  ``On first order formalism in string theory,''
  Phys.\ Lett.\ B {\bf 633} (2006) 375
  [hep-th/0510065].
  %%CITATION = HEP-TH/0510065;%%


  %\cite{Hohm:2013jaa}
\bibitem{Hohm:2013jaa}
  O.~Hohm, W.~Siegel and B.~Zwiebach,
  ``Doubled $\alpha'$-geometry,''
  JHEP {\bf 1402} (2014) 065
  [arXiv:1306.2970 [hep-th]].
  %%CITATION = ARXIV:1306.2970;%%


%\cite{Evans:1989xq}
\bibitem{Evans:1989xq}
  M.~Evans and B.~A.~Ovrut,
  ``Deformations of Conformal Field Theories and Symmetries of the String,''
  Phys.\ Rev.\ D {\bf 41} (1990) 3149.
  %%CITATION = PHRVA,D41,3149;%%

%\cite{Siegel:1993th}\cite{Hull:2009zb}}
\bibitem{Siegel:1993th}
  W.~Siegel,
  ``Superspace duality in low-energy superstrings,''
  Phys.\ Rev.\ D {\bf 48} (1993) 2826
  [hep-th/9305073].
  %%CITATION = HEP-TH/9305073;%%
  %126 citations counted in INSPIRE as of 01 Jul 2014

%\cite{Hull:2009zb}
\bibitem{Hull:2009zb}
  C.~Hull and B.~Zwiebach,
  ``The Gauge algebra of double field theory and Courant brackets,''
  JHEP {\bf 0909} (2009) 090
  [arXiv:0908.1792 [hep-th]].
  %%CITATION = ARXIV:0908.1792;%%
  %116 citations counted in INSPIRE as of 01 Jul 2014

%\cite{Lian:1992mn}
\bibitem{Lian:1992mn}
  B.~H.~Lian and G.~J.~Zuckerman,
  ``New perspectives on the BRST algebraic structure of string theory,''
  Commun.\ Math.\ Phys.\  {\bf 154} (1993) 613
  [hep-th/9211072].
  %%CITATION = HEP-TH/9211072;%%



%\cite{Kac:1996wd}
\bibitem{Kac:1996wd}
  V.~Kac,
 ``Vertex algebras for beginners,''
  Providence, USA: AMS (1996) 141 p. (University lectures series. 10)


%\cite{Zeitlin:2009hc}
\bibitem{Zeitlin:2009hc}
  A.~M.~Zeitlin,
  ``Beta-gamma systems and the deformations of the BRST operator,''
  J.\ Phys.\ A {\bf 42} (2009) 355401
  [arXiv:0904.2234 [hep-th]].
  %%CITATION = ARXIV:0904.2234;%%


%\cite{Zeitlin:2014xma}
\bibitem{Zeitlin:2014xma}
  A.~M.~Zeitlin,
  ``Beltrami-Courant Differentials and $G_{\infty}$-algebras,''
  arXiv:1404.3069 [math.QA].
  %%CITATION = ARXIV:1404.3069;%%


%\cite{Sen:1990hh}
\bibitem{Sen:1990hh}
  A.~Sen,
  ``On the Background Independence of String Field Theory,''
  Nucl.\ Phys.\ B {\bf 345} (1990) 551.
  %%CITATION = NUPHA,B345,551;%%


%\cite{Campbell:1990dz}
\bibitem{Campbell:1990dz}
  M.~Campbell, P.~C.~Nelson and E.~Wong,
  ``Stress tensor perturbations in conformal field theory,''
  Int.\ J.\ Mod.\ Phys.\ A {\bf 6} (1991) 4909.
  %%CITATION = IMPAE,A6,4909;%%


%\cite{Ovrut:1990ue}\cite{Ranganathan:1992nb}%\cite{Ranganathan:1993vj}%\cite{Cederwall:1995nc}
\bibitem{Ovrut:1990ue}
  B.~A.~Ovrut and S.~Kalyana Rama,
  ``Deformations of Superconformal Field Theories and Target Space Symmetries,''
  Phys.\ Rev.\ D {\bf 45} (1992) 550.
  %%CITATION = PHRVA,D45,550;%%


%\cite{Ranganathan:1992nb}%\cite{Ranganathan:1993vj}
\bibitem{Ranganathan:1992nb}
  K.~Ranganathan,
``Nearby CFTs in the operator formalism: The Role of a connection,''
  Nucl.\ Phys.\ B {\bf 408} (1993) 180
  [hep-th/9210090].
  %%CITATION = HEP-TH/9210090;%%


%\cite{Ranganathan:1993vj}
\bibitem{Ranganathan:1993vj}
  K.~Ranganathan, H.~Sonoda and B.~Zwiebach,
``Connections on the state space over conformal field theories,''
  Nucl.\ Phys.\ B {\bf 414} (1994) 405
  [hep-th/9304053].
  %%CITATION = HEP-TH/9304053;%%


%\cite{Cederwall:1995nc}
\bibitem{Cederwall:1995nc}
  M.~Cederwall, A.~von Gussich and P.~Sundell,
``Deformations in closed string theory: Canonical formulation and regularization,''
  Nucl.\ Phys.\ B {\bf 464} (1996) 117
  [hep-th/9504112].
  %%CITATION = HEP-TH/9504112;%%


%\cite{Hohm:2014eba}
\bibitem{Hohm:2014eba}
  O.~Hohm and B.~Zwiebach,
  ``Green-Schwarz mechanism and $\alpha'$-deformed Courant brackets,''
  arXiv:1407.0708 [hep-th].
  %%CITATION = ARXIV:1407.0708;%%


%\cite{Polchinski:1998rq}
\bibitem{Pol1}
  J.~Polchinski,
  ``String theory. Vol. 1: An introduction to the bosonic string,''
  Cambridge, UK: Univ. Pr. (1998) 402 p


%\cite{Giveon:1994fu}
\bibitem{Giveon:1994fu}
  A.~Giveon, M.~Porrati and E.~Rabinovici,
  ``Target space duality in string theory,''
  Phys.\ Rept.\  {\bf 244} (1994) 77
  [hep-th/9401139].
  %%CITATION = HEP-TH/9401139;%%



\end{thebibliography}
\end{document}